\documentclass[11pt]{article}
\usepackage{jheppub}

\pdfoutput=1
\usepackage{amsmath,amssymb}
\usepackage{epsfig}
\usepackage{hyperref}
\usepackage{color}
\usepackage{slashed}
\usepackage{placeins}

\usepackage{amsfonts,dsfont}
\usepackage{graphicx, rotating}
\usepackage{epstopdf}
\usepackage{latexsym}
\usepackage{rotating}

\usepackage[normalem]{ulem}

\usepackage{subfig}

\usepackage[export]{adjustbox}

\usepackage[shortlabels]{enumitem}

\usepackage[font={small}]{caption}   

\definecolor{myred}{rgb}{0.7, 0, 0}
\definecolor{myblue}{rgb}{0, 0, 0.7}
\definecolor{mygreen}{rgb}{0.04, 0.7, 0.5}

\hypersetup{colorlinks,citecolor=myred,linkcolor=myblue,urlcolor=myblue,linktocpage=true}

 \def\be   {\begin{equation}}   \def\ee   {\end{equation}}
 \def\ba   {\begin{array}}      \def\ea   {\end{array}}
 \def\bea  {\begin{eqnarray}}   \def\eea  {\end{eqnarray}}
 \def\bean {\begin{eqnarray*}}  \def\eean {\end{eqnarray*}}
 
 \def\bry{\begin{array}}
 \def\ery{\end{array}}

\numberwithin{equation}{section}

\begin{document}

\begin{flushright}
\footnotesize
TUM-HEP 1597/26 \\
\end{flushright}
\color{black}

\title{
Inverse Electroweak Baryogenesis
}

\author{Jacopo Azzola,}

\author{Oleksii Matsedonskyi,}

\author{Andreas Weiler}

\affiliation{Technische Universit\"at M\"unchen, Physik-Department, James-Franck-Strasse 1, 85748 Garching, Germany}

\emailAdd{jacopo.azzola@tum.de, oleksii.matsedonskyi@tum.de, andreas.weiler@tum.de}

\abstract{
We propose a mechanism for baryogenesis in which the baryon asymmetry is generated as an \emph{equilibrium response} of weak sphalerons in a region where electroweak sphaleron transitions remain unsuppressed, \mbox{$h/T\lesssim 1$}. A nonzero equilibrium baryon density arises in the presence of an approximately conserved global charge, carried either by new states with nonzero hypercharge, or by Standard Model fields themselves.
The required global charge asymmetry is generated during a phase transition that changes the strength of electroweak symmetry breaking, but need not coincide with the final electroweak phase transition. In particular, the mechanism can operate during an inverse electroweak phase transition, where baryon number is produced behind the advancing wall, in contrast to conventional electroweak baryogenesis. Because baryon production is decoupled from a direct first-order electroweak phase transition, the scenario can be realized at parametrically higher temperatures than standard electroweak baryogenesis, thereby weakening current experimental constraints. This framework provides a qualitatively distinct route to electroweak baryogenesis, with different parametric dependence, phase-transition dynamics, and phenomenological signatures.
}

\maketitle


\section{Introduction}

Electroweak baryogenesis (EWBG)~\cite{Shaposhnikov:1987tw,Cohen:1990it} is one of the most extensively studied scenarios for explaining the observed matter-antimatter asymmetry. Its defining feature is the reliance on high-temperature electroweak (EW) symmetry restoration followed by EW symmetry breaking during a first-order EW phase transition (EWPT)\footnote{See~\cite{Morrissey:2012db,Cline:2006ts} for reviews and~\cite{Bhusal:2025lvm} for recent alternative realizations.}. A first-order EWPT is a relatively easy-to-achieve modification of the Standard Model (SM) electroweak thermal history, but it is by no means the only possibility. A number of alternative EW phase transition scenarios have been studied in the literature~\cite{Shakya:2025mdh,Espinosa:2004pn,DiLuzio:2019wsw,Glioti:2018roy,Matsedonskyi:2022btb,Baldes:2018nel,Matsedonskyi:2020kuy,Matsedonskyi:2020mlz,Badziak:2025fdp,Biekotter:2022kgf,Inoue:2015pza,Biekotter:2021ysx,Aoki:2023lbz,Ai:2025vfi,Barni:2025ced,Barni:2024lkj,Badziak:2022ltm,Azzola:2026cwa}.
Furthermore, phase-transition-like behaviour can be realized locally in the vicinity of topological defects~\cite{Brandenberger:1994mq,Abel:1995uc,Brandenberger:2005bx,Bai:2021xyf,Schroder:2024gsi,Azzola:2024pzq,Sassi:2024cyb}. 
In this work we will mainly discuss a scenario allowing to create a baryon asymmetry after inverse EW phase transition, although we will also clarify how this can be generalized. 
Overall, we refer to the scenario as inverse electroweak baryogenesis in the sense that baryon number is generated in the region \emph{behind} the wall of a first-order phase transition, that modifies the strength of electroweak symmetry breaking. This is opposite to the standard EWBG where the weak sphalerons are active in front of the wall.

Throughout this work, we use the term \emph{first-order inverse EW phase transition} for any phenomenon that leads to walls having broken EW phase in front and unbroken behind them.
To provide a concrete framework for the dynamics underlying an inverse electroweak phase transition, we can consider the suitably re-shaped singlet-extended Standard Model introduced in Ref.~\cite{Azzola:2024pzq}. In the presence of an approximate $Z_2$ symmetry, the scalar potential admits two quasi-degenerate minima at $S=\pm v_S$. The vacuum manifold may therefore consist of spatial domains occupying these distinct minima, separated by domain walls that interpolate between $\pm v_S$.
In the presence of a positive Higgs–singlet portal interaction, $|H|^2 S^2$, the effective Higgs mass parameter decreases as $S$ approaches zero within the wall core. This can induce a wall-localized Higgs vacuum expectation value (VEV).
Consequently, at temperatures above $\sim 160$~GeV, the electroweak symmetry can be restored in the bulk plasma while being spontaneously broken inside the domain walls.
As a domain wall sweeps across a given spatial region, that region undergoes a sequence of transitions: from the unbroken to the broken electroweak phase upon entering the wall core, and subsequently back to the unbroken phase once the wall has passed. This realizes the desired local inverse EWPT.

While a first-order phase transition provides the required departure from equilibrium~\cite{Sakharov:1967dj}, an efficient source of baryon number is still needed. We show, in agreement with Refs.~\cite{Cline:1993vv,Antaramian:1993nt,Dick:1999je}, that unsuppressed weak sphalerons operating behind the phase-transition wall can generate a nonzero equilibrium baryon density in the presence of suitable approximate global charges. We demonstrate that such charges can arise during the phase transition from CP violation associated with a space-dependent complex phase of the Standard Model top-quark mass, a source commonly employed in conventional electroweak baryogenesis~\cite{Joyce:1994fu,Joyce:1994zt,Cline:2000kb}. As an explicit illustration, we present a preliminary study of a concrete channel that produces a global charge asymmetry carried by an inert Higgs doublet.

\begin{figure}[t]
\center
\includegraphics[width=0.5 \textwidth]{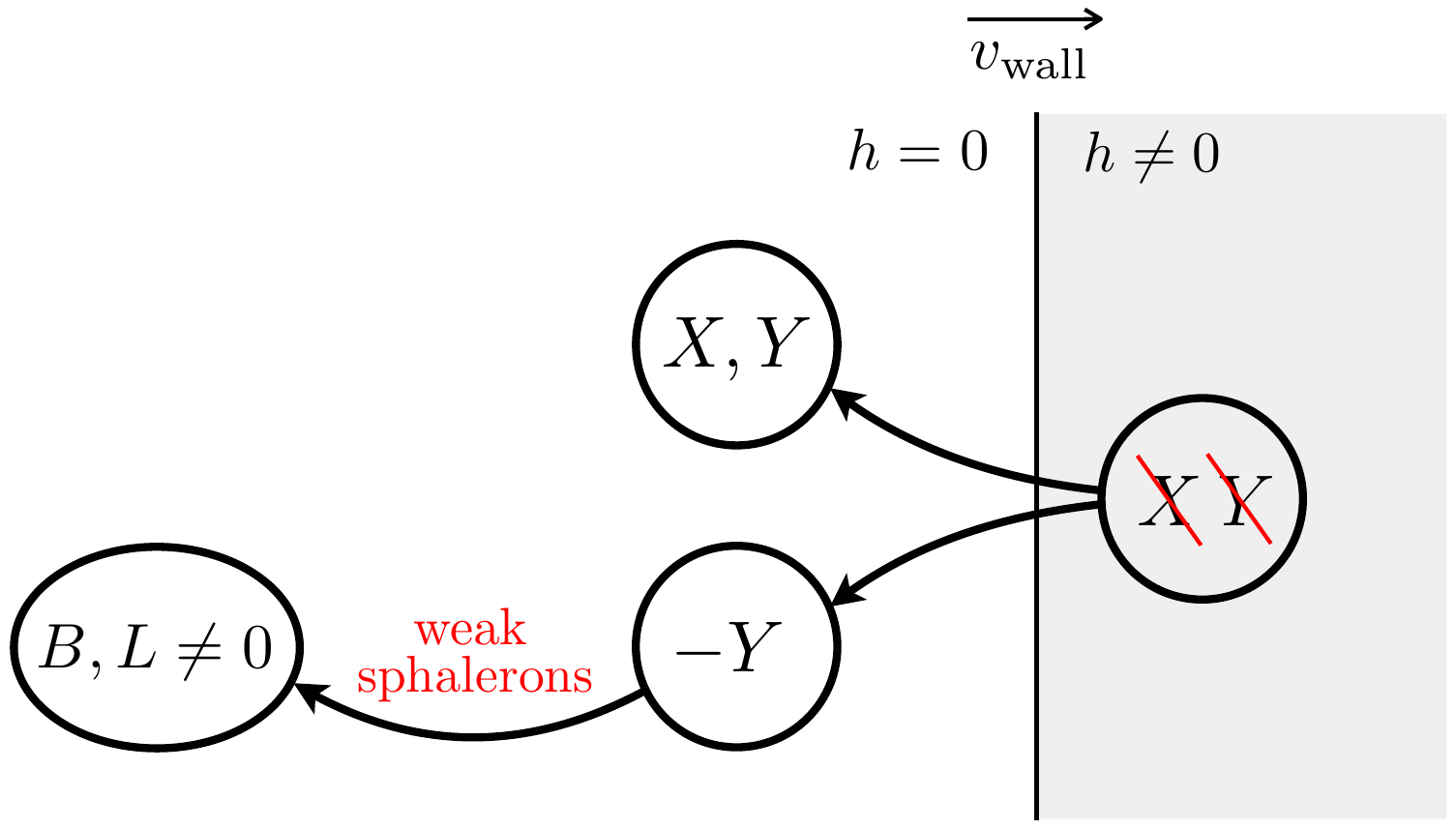} \hspace{0.5cm}
\caption{
Schematic illustration of the mechanism of baryon-number generation during an inverse electroweak phase transition. CP-violating interactions with the phase boundary, together with $X$-violating processes in front of the wall, generate a static excess of an approximately conserved global charge $X$ in the region behind the wall. Since the $X$-charge carriers also carry hypercharge, overall gauge-charge neutrality (enforced by plasma screening) requires compensating hypercharge distributed   among the Standard Model species. The resulting nonzero hypercharge of SM induces a nonvanishing equilibrium baryon number once weak sphaleron processes become active.}
\label{fig:sketch}
\end{figure}

A schematic illustration of the mechanism is shown in Fig.\ref{fig:sketch}, assuming some new global charge $X$.
The essential logical steps of the mechanism are as follows:
\begin{itemize}
\item
A first-order phase transition, together with CP violation and $X$-charge violation localized at the inverse EWPT wall, generates a net global $X$ asymmetry. We assume that the $X$-charge carriers also carry non-zero hypercharge.

\item
The generated $X$ asymmetry is conserved in the plasma behind the phase transition wall. Since the $X$-charge carriers are hypercharged, and overall hypercharge neutrality is dynamically enforced, the Standard Model plasma develops a compensating non-zero hypercharge density.

\item
In the presence of a non-zero hypercharge density in the Standard Model plasma, weak sphaleron processes are biased toward an equilibrium configuration with non-vanishing $B+L$. An analogous effect occurs in the electroweak-broken phase for $h/T<1$, where a non-zero electric charge density in the plasma similarly biases sphaleron transitions.

\item
The $X$-charge carriers must remain stable until temperatures where weak sphaleron processes fall out of equilibrium, and the generated baryon asymmetry becomes frozen in. Assuming standard model-like thermal history, this means that the $X$-charge carriers should survive till $T \lesssim 130$~GeV. 
\end{itemize}

In the following, we analyze these ingredients in detail. We also discuss variants in which the approximate global charge is carried by Standard Model fields, as well as broader generalizations of the mechanism.

\section{Equilibrium Baryon Number}
\label{s:chieq}

Let us begin by discussing the role of conserved charges in the EW-symmetric phase. As shown, for example, in Refs.~\cite{Cline:1993vv,Antaramian:1993nt}, the presence of approximately conserved charges associated with right-handed electrons, or with a new $U(1)$ symmetry overlapping with hypercharge, can protect the baryon number from washout by weak sphalerons. We will re-derive analogous arguments in the context of our setup and show how a non-zero equilibrium baryon number can arise in the presence of such charges.

\subsection{Warm-up}

As a warm-up, let us simply restate the main line of arguments of Ref.~\cite{Antaramian:1993nt}, keeping details and refinements for later. We use the linear-response relation for the charge density $\delta n_i=\chi_{i}\mu_i$ valid for small $\mu_i/T$ (see Appendix~\ref{s:appendix}), and define the $\chi$-weighted inner product
\be
(A,B)\equiv A^T.\chi.B=\sum_i A_i\,\chi_i\,B_i,
\ee
so that the total $A$-charge density stored in chemical potentials $\mu$ is given by $(A,\mu)$. When the indices of $\chi$ are not explicitly stated, it is assumed to be a diagonal matrix $\chi_i \delta_{ij}$.

\begin{enumerate}

\item
In equilibrium, the vector of chemical potentials can be decomposed in the conserved charges $Q$. Assuming only two such charges --- hypercharge and a new $U(1)_X$ --- we have 
\be
\mu_i = c_Y (Q_Y)_i + c_X (Q_X)_i.
\ee

\item
Macroscopic gauge-charge densities are efficiently neutralized in a relativistic plasma due to Debye screening. 
Consequently, on length scales much larger than the screening length $\lambda_D\sim (gT)^{-1}$, the total hypercharge density vanishes in the EW-symmetric phase,
\be
0 = (Q_Y,\mu) = c_Y\, (Q_Y,Q_Y) + c_X\, (Q_Y,Q_X).
\ee
Therefore
\be\label{eq:toycYcX}
c_Y = - c_X \frac{(Q_Y,Q_X)}{(Q_Y,Q_Y)}.
\ee
Hence the coefficients $c_X$ and $c_Y$ are linearly-dependent as long as the hypercharge overlaps with the $X$-charge, $(Q_Y,Q_X) \ne0$.

\item 
At the same time, a non-vanishing $X$-charge density  implies
\be
X \equiv (Q_X,\mu) = c_Y \, (Q_X,Q_Y) + c_X \, (Q_X,Q_X) \ne 0.
\ee
Combining with Eq.~\eqref{eq:toycYcX} we obtain
\be\label{eq:cYsimp}
c_Y = \frac{(Q_Y,Q_X)}{(Q_Y,Q_X)^2 - (Q_X,Q_X) \, (Q_Y,Q_Y)} X \ne 0.
\ee
These constraints are central to our mechanism: if an approximately conserved charge $X$ is carried by hypercharged (or electrically charged) states, the neutrality condition forces compensating chemical potentials among SM species, which biases sphaleron-mediated equilibration toward a nonzero baryon number as we see in the next step.

\item
Finally, assuming that the $X$-charge carriers are not baryons, $(Q_B,Q_X)=0$, the overall baryon number density is given by
\be
(Q_B,\mu) = 
c_Y \, (Q_B,Q_Y),
\ee
with $(Q_B,Q_Y)\ne0$ in the SM. Using the explicit form of $c_Y$ given in Eq.~\eqref{eq:cYsimp} we then find that 
\be
(Q_B,\mu) 
\, \propto \, (Q_Y,Q_X) \, X,
\ee
i.e. the baryon number is non-zero in the presence of the $X$-charge density. 

\end{enumerate}

\subsection{One SM generation}

After sketching the core argument, let us derive it more carefully, including for now one full SM fermion generation, and accounting for all the relevant conserved charges.

To describe the evolution of the particle asymmetries $\delta n_i$ generated after the wall has passed\footnote{We discuss the concrete $\delta n_i$ production channels in the next section.} and the associated velocity perturbations have dissipated, we employ the corresponding homogeneous-space Boltzmann equation
\be
\dot{\delta n}_i = {\cal C}, 
\ee
where ${\cal C}$ stands for collision terms. Assuming thermal equilibrium 
\be
n \propto  \int d^3 p \, \left(\exp[(E-\mu)/T]\pm 1\right)^{-1},
\ee
with small asymmetries $\mu \ll T$, such that $\delta n \propto \mu$, one arrives at~\cite{Cline:2020jre} 
\be\label{eq:muevolfs}
\chi_{i j} \dot \mu_j
= -(\Gamma_{\mu})_{i j} \mu_j, 
\ee
We neglect the small $\partial_t T$ in Eq.~\eqref{eq:muevolfs} and below. The matrix $\Gamma_{\mu}$ originating from the collision terms describes species-changing interactions.   
Eq.~\eqref{eq:muevolfs} can be integrated to obtain 
\bea \label{eq:mueq0}
\mu(t) &=& \exp\left[-\chi^{-1}.\Gamma_{\mu} \,t \right].\mu_S,
\eea
where $\mu_S$ is the initial perturbation that was produced by the phase transition wall. Let us now assume that, at some time $t_{eq}$, a subset of interactions has reached equilibrium, while the remaining ones are too weak to significantly affect the chemical potentials and can thus be neglected in $\Gamma_\mu$. The components of $\mu_S$ along the non-zero eigenvectors of $\chi^{-1}\Gamma_{\mu}$ are exponentially damped and can therefore be ignored. The equilibrium distribution at $t_{eq}$ is then approximately given by
\bea \label{eq:mueq}
\mu &\simeq&  \hat P_0.\mu_S,
\eea 
where $\hat P_0$ is a projector on the conserved (zero eigenvector) subspace of the matrix $\chi^{-1}.\Gamma_{\mu}$.
The basis of the zero-subspace can be chosen as a set of conserved linearly-independent charges $Q$. To see this, first note that the zero-eigenvectors $v_0$ satisfy $\Gamma_\mu.v_0 = 0$, and hence by $\Gamma$ matrix symmetry $v_0^T.\Gamma_\mu=0$. Using this fact, and contracting Eq.~\eqref{eq:muevolfs} with $v_0$ one finds
\be
\partial_t (v_0^T.\chi.\mu) = - v_0^T.\Gamma_\mu.\mu = 0,
\ee
which is a definition of a conserved charge $Q^T.\delta n$, with $Q=v_0$. Hence we obtain a decomposition
\bea
\mu &\simeq& \hat P_0.\mu_S = \sum_\alpha c_\alpha \, Q_\alpha, 
\eea
where $\alpha$ runs over linearly-independent conserved charges.

We will now use this $\mu$ decomposition to derive some important general properties regarding the baryon number, first assuming only one SM fermion family for simplicity. The vector of chemical potentials is given by: 
\be
\mu=(\mu_{u_L},\mu_{u_R},\mu_{d_L},\mu_{d_R},\mu_{e_L},\mu_{e_R},\mu_{\nu_L},\mu_{\nu_R},\mu_{H_0},\mu_{H_+},\mu_{T_3}, \dots),
\ee
where $\dots$ stand for the chemical potentials of $X$-charge carriers that we specify later.
We can then decompose $\mu$ as: 

\be\label{eq:mudec2}
\mu = c_Y Q_Y + c_{B-L} Q_{B-L} + c_{\nu_R} Q_{\nu_R} + c_X Q_X + c_{T_3} Q_{T_3}.
\ee
At large enough times $t T \gtrsim 10^6$ the weak sphaleron processes reach equilibrium, so the individual baryon and lepton number symmetries are reduced to $B-L$. We also include a $U(1)$ symmetry corresponding to the conservation of the right-handed neutrino number~\footnote{We assume Dirac neutrinos, although we expect our results to extend to scenarios with Majorana masses, provided the corresponding lepton-number-violating effects are sufficiently decoupled.}. The corresponding Yukawa interaction $y_\nu \bar l_L \tilde H \nu_R $ is too weak and never equilibrates at the relevant time scales. All other Yukawa interactions are assumed to be in equilibrium. We will also postulate an additional $U(1)_X$ symmetry. We further assume that the $X$ charge is carried by a BSM state without baryon and lepton charge. Finally, we included a charge vector for the weak isospin $T_3$.

First, imposing vanishing $B-L$~\footnote{See Ref.~\cite{Harvey:1990qw} for the analysis in the presence of a non-vanishing $B-L$.} we find
\bea
0 = (Q_{B-L}, \mu) 
&=& 
c_Y (Q_{B-L},Q_Y) + c_{B-L} (Q_{B-L},Q_{B-L}) + c_{\nu_R} (Q_{\nu R},Q_{B-L})\nonumber \\
&=&
4 c_Y (Q_{B},Q_Y) + 4 c_{B-L} (Q_{B},Q_{B}) - c_{\nu_R} (Q_{\nu_R},Q_{\nu_R}), \label{e:QBmL}
\eea
where we used the identities
\bea
&(Q_{B-L},Q_Y) = 4 (Q_{B},Q_Y), \;\;\;
(Q_{B-L},Q_{B-L}) = 4 (Q_{B},Q_{B}),& \\
&(Q_{\nu_R},Q_{B-L})=-(Q_{\nu_R},Q_{\nu_R}),  \;\;\;
(Q_{B-L},Q_X)=0.&
\eea
We can now evaluate the baryon charge of the equilibrium $\mu$ distribution
\bea\label{e:cnuR}
(Q_{B},\mu) 
&=& 
c_Y (Q_{B},Q_Y) + c_{B-L} (Q_{B},Q_{B-L})\\
&=&
c_Y (Q_{B},Q_Y) + c_{B-L} (Q_{B},Q_{B}) \\
&=&
\frac 1 4 c_{\nu_R} (Q_{\nu_R},Q_{\nu_R}),
\eea
where in the last step we used Eq.~\eqref{e:QBmL}. Hence in the absence of the conserved charge associated with the right-handed neutrino, $c_{\nu_R}=0$, the baryon number would vanish. If instead $Q_{\nu_R}$ is conserved, but no $\nu_R$ asymmetry is present, $c_{\nu_R}$ is generally non-zero and can be fixed from:
\be\label{eq:nonuR}
(Q_{\nu_R},\mu) = 0 \;\; \Rightarrow \;\; c_{\nu_R} = c_{B-L}.
\ee
The decoupling of the neutrinos is necessary, but not a sufficient condition for a non-zero $B$. Let us demonstrate the necessity of the $U(1)_X$ symmetry. 
Imposing the global hypercharge neutrality 
\bea
0 = (Q_{Y},\mu) 
&=& 
c_Y (Q_{Y},Q_Y) + c_{B-L} (Q_{Y},Q_{B-L}) + c_{X} (Q_{Y},Q_{X})\nonumber \\
&=&
c_Y (Q_{Y},Q_Y) + 4 c_{B-L} (Q_{Y},Q_{B}) + c_{X} (Q_{Y},Q_{X}), \label{e:QY}
\eea
together with Eqs.~\eqref{e:QBmL},~\eqref{e:cnuR},~\eqref{eq:nonuR} we find the baryonic charge to be
\bea
(Q_{B},\mu) 
&=&
\frac{(Q_Y,Q_B)(Q_{\nu_R},Q_{\nu_R})(Q_Y,Q_X)}{4(Q_B,Q_B)(Q_Y,Q_Y)-(Q_Y,Q_Y) (Q_{\nu_R},Q_{\nu_R}) -16(Q_Y,Q_B)^2} c_X \nonumber \\
&+&
\frac{ (Q_Y,Q_Y)(Q_B,Q_B) - 4 (Q_Y,Q_B)^2}{4(Q_B,Q_B)(Q_Y,Q_Y)-(Q_Y,Q_Y) (Q_{\nu_R},Q_{\nu_R})-16(Q_Y,Q_B)^2} (Q_{\nu_R},\mu). 
\eea
Hence one option allowing for a non-zero equilibrium baryon number is a presence of a conserved $X$-charge, $c_X\ne 0$, that has an overlap with a hypercharge, $(Q_Y,Q_X)\ne 0$.
Another possibility is instead to have a non-vanishing number of right-handed-neutrinos from the start, $(Q_{\nu_R},\mu) \ne 0$, in agreement with the earlier results of Ref.~\cite{Dick:1999je}.

Note that $c_{T_3}$ in the $\mu$ decomposition~\eqref{eq:mudec2} wouldn't contribute to any of the charges discussed above because of the $Q_{T_3}$ orthogonality to all other charges. 
Furthermore, the condition $(Q_{T_3},\mu)=0$ simply implies $c_{T_3}=0$~\cite{Antaramian:1993nt} and $\mu_{T_3}=0$~\cite{Harvey:1990qw}.

One subtlety omitted in the above derivation is that we neglected the potential presence of approximately conserved charges associated with the right-handed charged leptons. While they typically can be considered in equilibrium at temperatures $T\sim 100$~GeV, their finite equilibration time can be important for small-size topological defects, if the EW symmetry restoration is confined to them. For example, a domain wall containing a region of restored EW symmetry, propagating through space that is in the broken phase, induces a local inverse EW phase transition at its leading edge~\cite{Azzola:2024pzq}. The region is subsequently returned to the broken phase on a timescale determined by the wall thickness. 
If this timescale is short enough, an additional $U(1)_{l_R}$ should be accounted for. Adding the corresponding charge vector into the $\mu$ decomposition~\eqref{eq:mudec2}, and using the constraints from $(Q_{B-L},\mu)=0$, and $(Q_{Y},\mu)=0$ one finds, independently on the presence of the $U(1)_X$, the following baryon charge: 
\be
(Q_{B},\mu) = \frac{(Q_B,Q_B)}{4 (Q_B,Q_B) - 2 (Q_{\nu_R},Q_{\nu_R})} \big\{(Q_{l_R},\mu) + (Q_{\nu_R},\mu) \big\},
\ee
where we imposed the SM identity $(Q_B,Q_B)=2(Q_B,Q_Y)$.
Hence, on one hand if the $l_R$ is decoupled and $(Q_{l_R},\mu)=(Q_{\nu_R},\mu)=0$, the baryon charge is zero even in the presence of the $X$ charge. On the other, in the absence of $X$ charge the baryon density can be non-zero as long as there is a non-zero $(Q_{l_R},\mu)$ or $(Q_{\nu_R},\mu)$. 

In summary, depending on the time elapsed since the phase transition relative to the weak sphaleron rate $\Gamma_{\text{ws}}$ and the charged-lepton Yukawa equilibration rate $\Gamma_{y_{l}}$, the system can fall into one of the following regimes:
\begin{itemize}
\item
$\Gamma_{\text{ws}}^{-1} < t < \Gamma_{y_{l}}^{-1}$:
non-zero equilibrium baryon number requires a non-zero right-handed lepton charges $\mu_{l_R}$.  
Note however that in this case the asymmetry would be washed out at $t > \Gamma_{y_{l}}^{-1}$. To prevent this sphaleron washout, the EW symmetry should get broken again rather quickly. 
Such short EW restoration periods can take place for example in domain walls~\cite{Brandenberger:1994mq, Azzola:2024pzq, Sassi:2024cyb}. In the following we will analyze the case with a non-zero $\tau_R$ asymmetry generated by the phase transition.
\item
$\Gamma_{\text{ws}}^{-1}, \Gamma_{y_{l}}^{-1} < t $:
non-zero equilibrium baryon number can be provided by a non-zero $X$ charge density. As an example of such an additional charge we will employ a $U(1)$ charge associated with an inert Higgs doublet $\Phi=\{\Phi_+, \Phi_0\}^T$. In this case there is no relevant limits on the duration of the EW-symmetric phase.  
\item
Finally, a right-handed-neutrino asymmetry can act as an effectively conserved charge and thereby induce a non-zero baryon number, provided the right-handed neutrinos remain decoupled from the thermal plasma. For the observed small neutrino masses, this condition is typically satisfied throughout the epoch in which weak sphalerons are active.
\end{itemize}

\begin{figure}[t]
\center
\includegraphics[width=0.59 \textwidth]{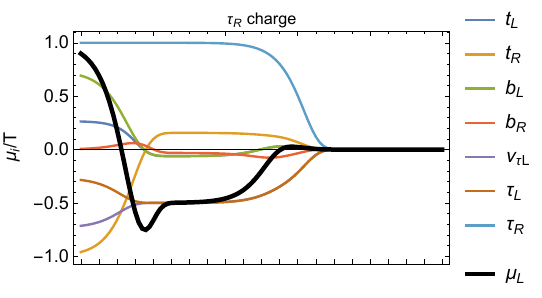}\\
\vspace{-.5cm}
\hspace{-1.07cm}\includegraphics[width=0.463\textwidth]{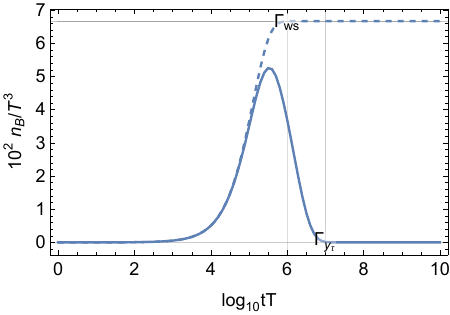}
\caption{{\bf Upper plot:}
Time evolution of some of the chemical potentials, and the combined left-handed chemical potential $\mu_L$ (in black) for a unit initial $\mu_{\tau_R}/T$. {\bf Lower plot:} Time evolution of the baryon number density for the same case. Gray vertical lines labeled with $\Gamma_{\text{ws}}$ and $\Gamma_{y_\tau}$ show the times at which weak sphalerons and tau-Yukawa-mediated interactions equilibrate. Blue dashed line corresponds to the evolution with all lepton Yukawas set to zero, demonstrating the effect of $n_{l_R}$ charges if they were exact.}
\label{fig:evolnum1}
\end{figure}

\begin{figure}[t]
\center
\includegraphics[width=0.59 \textwidth]{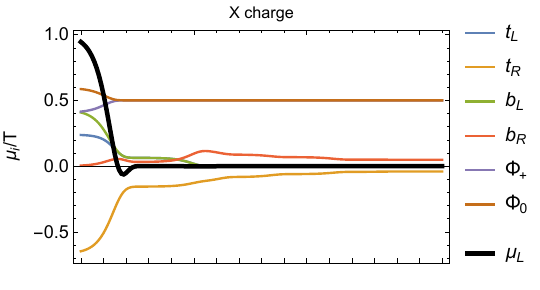}\\
\vspace{-.25cm}
\hspace{-1.07cm}\includegraphics[width=0.463\textwidth]{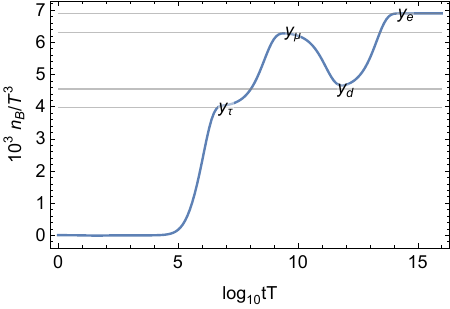}
\caption{Same as in Fig.\ref{fig:evolnum1} but for the unit initial $\mu_X/T = (\mu_{\Phi_+}+\mu_{\Phi_0})/T$ and vanishing $\mu_{\tau_R}/T$. Gray horizontal lines in the lower  plot correspond to the equilibrium baryon number values reached upon equilibration of corresponding Yukawa interactions.}
\label{fig:evolnum2}
\end{figure}

\subsection{Full SM}

Qualitatively, these conclusions also apply to the full system of SM particles. To analyze it, we use a more straightforward procedure based on the explicit expression for the projection operator $\hat P_0$. To model the initial $\mu$ distribution produced by the phase transition wall, we first take a generic vector of chemical potentials $\mu_S$ and project out those charges that are not generated in sizeable amounts during the inverse EWPT at $t = t_i$, using the operator $\mathds{1} - \hat P_0(t_i)$. 

Let us identify these charges first for the case without the $U(1)_X$ and $\mu_{\nu_R}$, and assuming that a phase transition wall has produced some initial amount of $\tau_R$ asymmetry. Starting from the maximal global symmetry of the SM, $U(3)^6 \times SU(2) \times U(1)$, which corresponds to 20 conserved quantities, we descend to 18 once we account for the fast $t_R$-number violation in the vicinity of the phase-transition wall, and the previously assumed $\tau_R$-number violation. 
The corresponding conserved charges, in obvious notations, are
\be
Q,\, Y,\, B_{1,2,3},\, L_{1,2,3},\, n_{q'_{Ri}},\, n_{l'_R},\, n_{\nu_{lR}},
\ee
where $q'_{Ri}$ are right-handed quarks except for the top, $l'_R$ are right-handed charged leptons except for the tau, $\nu_{lR}$ are all right-handed neutrinos, and we assume no flavour mixing for simplicity.

We then project the resulting vector on the conserved charges at a given equilibrium time $t_{eq}$ before the $\tau_R$ equilibration, $\Gamma_{\text{ws}}^{-1} < t_{eq} < \Gamma_{y_{\tau}}^{-1}$, to find the corresponding equilibrium distribution. At this point the conserved charges can be chosen as\footnote{Right-handed quark differences appear in the list since they are preserved by strong sphalerons, while corresponding Yukawas are not efficient at that point.}  
\be
Q,\, Y,\, B_{i}-L_{i},\, n_{s_{R}}-n_{u_{R}},\, n_{u_{R}}-n_{d_{R}},\, n_{l'_R},\, n_{\nu_{lR}}.
\ee

To proceed, we should obtain an explicit expression for the projector on the conserved subspace defined by the charges $Q_\alpha$:
\be\label{eq:projop2}
\hat P_0.\mu = c_\alpha Q_\alpha.
\ee
Using the fact that the projected out part of $\mu$, i.e. $(\mathds{1}-\hat P_0).\mu$, is not conserved, and hence has to have vanishing charges $Q_\alpha$, we find that
\be
0 = Q_\alpha^T.\chi.[\mathds{1}-\hat P_0].\mu = Q_\alpha^T.\chi.\mu - c_\beta \, Q_\alpha^T.\chi.Q_\beta.
\ee
This allows to derive the decomposition coefficients as
\be
c_\alpha = ({\cal M}^{-1})_{\alpha \beta} \, (Q_\beta,\mu),
\ee
with $\beta$ summation understood and
\be
{\cal M}_{\alpha \beta} \equiv (Q_\alpha,Q_\beta).
\ee

Knowing how the conserved charges define the projector, we can immediately find the baryon number density contained in $\hat P_0(t_{eq}).[\mathds{1} - \hat P_0(t_{i})].\mu_S$: 
\be\label{eq:nBtau}
n_B^{(\tau)}/T^3 = \frac {1} {15} \mu_{\tau_R}/T,
\ee
where the baryon number density is defined as
\be
n_B = \frac 1 3 \sum n_q = \frac {N_c} 3 \sum_{q=\{q_{iL},q_{iR}\}} \frac{\mu_q T^2}{6},
\ee
and $\mu_{\tau_R}$ is the $\tau_R$-component of the initial $\mu$ distribution.

Repeating the same procedure in the presence of the $X$-charge of an inert Higgs doublet (assuming it's sourced at the time of inverse EWPT $t_i$, but conserved later on), and neglecting a possible $\tau_R$ asymmetry, one finds that the maximal amount of baryon number density is produced after all the right-handed charged leptons equilibrate, and the following charges are preserved:
\be
Q,\, Y,\, B-L,\, n_{\nu_{lR}},
\ee
for which one gets
\be\label{eq:nBX}
n_B^{(X)}/T^3 = \frac {1}{145} \mu_{X}/T,
\ee
with $\mu_{X}$ being the sum of the $X$-charged particles' chemical potentials.

In Fig.\ref{fig:evolnum1} and Fig.\ref{fig:evolnum2} we show the results of the numerical solution of the evolution equation~\eqref{eq:mueq0}, in the presence of $\tau_R$ and $X$ charges respectively, using for $\Gamma$ the transition rates provided in Ref.~\cite{Cline:2021dkf}~\footnote{Note that estimates provided by different groups give significantly different rates in some cases. See~\cite{Barni:2025ifb} for a recent re-evaluation of these parameters.}. We treat independently the pairs of complex components of the Higgs doublets, and we also include a chemical potential $\mu_{T_3}$ for $T_3$ to make sure all the charge conservations are respected. 
For the initial vector $\mu_S/T$ we take a uniform vector, normalized such that after the action of $\mathds{1} - \hat P_0(t_{i})$ one has $\mu_{\tau_R}/T=1$ and $\mu_{X}/T=(\mu_{\Phi_+}+\mu_{\Phi_0})/T=1$ respectively. While this specific choice of the initial $\mu$ distributions affects the details of the evolution, it does not affect the quoted equilibrium values, that are uniquely fixed by $\mu_{\tau_R}$ and $\mu_{X}$.

The next important question concerns the fate of the equilibrium baryon density after the unavoidable direct EW phase transition.
If this transition is SM-like, electroweak symmetry becomes weakly broken with $h/T<1$ starting at $T\simeq160~\text{GeV}$, and only enters the strongly broken regime $h/T>1$ around $T\simeq130~\text{GeV}$. 
Throughout this temperature interval, weak sphalerons are not sufficiently suppressed, while the EW symmetries are already broken. One may therefore worry that the previously generated baryon asymmetry is erased.
In principle, if the direct EWPT is sufficiently fast (for instance, a strongly first-order global phase transition, see e.g.~\cite{Li:2025kyo}), sphaleron processes would not have enough time to significantly affect the baryon number, irrespective of the conserved charges of the system. 
However, there exists an additional symmetry-based mechanism that can protect the baryon asymmetry even if the EWPT is slow.
To see this, note that the symmetry arguments presented at the beginning of this section never relied on the conservation of weak isospin, but only on hypercharge conservation. 
Consequently, the entire analysis can be repeated in the broken EW phase (at $h/T<1$, where only $B-L$ is conserved) by replacing hypercharge with the electric charge, which is preserved after EW symmetry breaking.

One then arrives at the conclusion that a non-zero baryon number survives as long as there exist a population of BSM particles carrying a conserved particle number $X$ and a non-zero electric charge. 
Such electrically charged states cannot be absolutely stable and must eventually decay, transferring their electric charge to SM particles. 
However, provided that this decay occurs after $T\simeq130~\text{GeV}$, when $h/T>1$ and sphaleron processes are sufficiently suppressed so that $B$ becomes an effectively conserved quantity, this mechanism successfully protects the baryon asymmetry from sphaleron wash-out.

To obtain the resulting baryon number in the case of the inert Higgs doublet, we project the previously determined chemical potential vector onto the conserved charges in the broken EW phase, obtaining\footnote{Note that with such a $B/X$ ratio, and the overall $n_B$ fixed to the observed amount, the contribution of the fermionic asymmetries to the hypercharge field evolution does not compete with the Hubble expansion rate and hence does not trigger the hypercharge field chiral instability~\cite{Bodeker:2019ajh}.}
\be\label{eq:nBXQ}
n_B^{(X)}/T^3 = \frac{1}{353}\,\mu_X/T.
\ee

We have therefore established the baryon asymmetry supported by conserved charges in the three relevant regimes: $\tau_R$ asymmetry in the unbroken phase, Eq.~\eqref{eq:nBtau}; $X$ asymmetry in the EW unbroken phase, Eq.~\eqref{eq:nBX}; and $X$ asymmetry in the broken EW phase, Eq.~\eqref{eq:nBXQ}. All three results are given per unit $\mu/T$ stored in the corresponding approximately conserved charge. We now turn to the production mechanisms for these charge asymmetries.

\section{Conserved Charges from Inverse EWPT}
\label{s:concur}

We now turn to the production of the approximately conserved charges required in the previous section. The basic idea is simple. A CP-violating phase-transition wall generates local top quark asymmetries in its vicinity. The $X$-charge-violating interactions in front of the wall will convert the top quark excess into an $X$-charge excess, while an opposite process will not happen behind the wall where the $X$-charge is preserved. Overall, the advancing wall then leaves behind a static trail of conserved $X$-charge. In our setup, such a trailing asymmetry can subsequently bias weak sphalerons toward a nonzero equilibrium baryon number.

\subsection{Fluid equations}

To describe this process quantitatively, we use the fluid-equation approach of Ref.~\cite{Cline:2020jre}, based on two moments of the Boltzmann equation. Besides the chemical potentials $\mu_i$, this formalism also tracks the velocity perturbations $u_i$ of the various species. In the wall frame, the transport equations take the form
\bea
-D_1 \,\mu'_i + u'_i + v_{\text{w}} \gamma_{\text{w}} (m^2)' Q_1 \,\mu_i &=& (S^o_{h 1})_i + (\delta C_1)_i, \label{eq:fluideq1} \\
-D_2 \,\mu'_i - v_{\text{w}} \,u'_i + v_{\text{w}} \gamma_{\text{w}} (m^2)' Q_2 \,\mu_i +  (m^2)' \bar R \,u_i &=& (S^o_{h 2})_i + (\delta C_2)_i. \label{eq:fluideq2}
\eea
Here primes denote derivatives with respect to the coordinate $z$ normal to the wall, and $v_{\text w}$ is the wall velocity. The functions $D_{1,2}$ and $Q_{1,2}$ are momentum-averaged coefficients defined in Ref.~\cite{Cline:2020jre}; their explicit form will not be important for the qualitative discussion below. The terms $S^o_{h l}$ are the CP-violating sources, while $(\delta C_l)\propto \mu,u$ denote collision terms, including both species-changing interactions and the damping of velocity perturbations.

Let us consider the first equation more closely. It can be rewritten as
\be\label{eq:fluid1}
\partial_z(|D_1| \,\mu_i+ u_i) = (S^o_{h 1})_i + (\delta C_1)_i,
\ee
and represents the particle number conservation equation in the wall frame. 
A useful analogy can be derived using a particle charge density $q(z)$, with a wall-frame equilibrium velocity $-v_{\text{w}}$ and the actual perturbed velocity $-v_{\text{w}} - \delta v(z)$. Corresponding current conservation equation reads
\bea\label{eq:cons1} 
\partial_z (v_{\text{w}} \,q  + \delta v  \,q) 
= S_q + \Gamma_q \, q, 
\eea
where the current non-conservation is caused by the source $S_q$ and the charge decays with a rate $\Gamma_q$. Noticing that $|D_1| \to v_{\text{w}}$ in the massless limit, one can see that $|D_1|\mu$ of Eq.~\eqref{eq:fluid1} matches onto $v_{\text{w}} q$ of Eq.~\eqref{eq:cons1}, and $u$ onto $\delta v \, q$.  

Now consider a conserved linear combination of charges, denoted by the subscript $c$. Away from the source and in the absence of charge-violating interactions, $(\delta C_1)_c=0$, Eq.~\eqref{eq:fluid1} implies
\be\label{eq:fluid1conssol}
|D_1| \,\mu_c = - u_c + const.
\ee
This suggests that for $u_c=0$ a $\mu_c=const$ solution behind the wall can exist.  
To determine whether such a constant trailing solution is physically realized, we examine the second transport equation. Substituting Eq.~\eqref{eq:fluid1conssol} into Eq.~\eqref{eq:fluideq2}, and taking the mass to be approximately constant, gives
\bea\label{eq:fluid12}
\left(\frac{D_2}{|D1|} - v_{\text{w}}\right) \,u_c' &=& (S^o_{h 2})_c + (\delta C_2)_c.
\eea
For conserved currents, the part of $(\delta C_2)_c$ proportional to $\mu_c$ vanishes, but elastic scattering still damps the velocity perturbation,
\be
(\delta C_2)_c \equiv -\Gamma_u\,u_c .
\ee
We then find that the free ($S=0$) solution of Eq.~\eqref{eq:fluid12} is of the form
\be\label{eq:uc}
u_c = u_0 \exp\left[-\frac{\Gamma_u}{D_2/|D1|-v_{\text{w}}} z\right], 
\ee
while $\mu$ can be obtained from Eq.~\eqref{eq:fluid1conssol}:
\be\label{eq:muc}
\mu_c =  - \frac{u_0}{|D1|} \exp\left[-\frac{\Gamma_u}{D_2/|D1|-v_{\text{w}}} z\right] + \mu_0.
\ee
Note that the argument of the exponents is always negative~\cite{Cline:2020jre}.

These solutions have a simple interpretation. The exponentially-decaying solution with $\mu_0=0$ and $u_0\neq 0$ has zero net current in the wall frame, $|D_1|\mu_c+u_c=0$, and describes a charge distribution moving together with the wall.
The solution with $\mu_0\neq 0$ and $u_0=0$ corresponds to a homogeneous and static conserved charge density in the plasma frame. Such a constant conserved charge density in front of the wall would be unphysical, since it would correspond to an asymmetry present already before the phase transition. The physically relevant constant asymmetries are instead those generated locally at the wall and deposited behind it. In conventional EWBG these are the conserved $B$ and $L$ asymmetries left behind in the broken phase. In the present setup, the same logic applies to conserved global charges deposited in the electroweak-symmetric phase behind an inverse wall.

\subsection{Point-like source}

\begin{figure}[t]
\center
{\includegraphics[width=0.38\textwidth]
{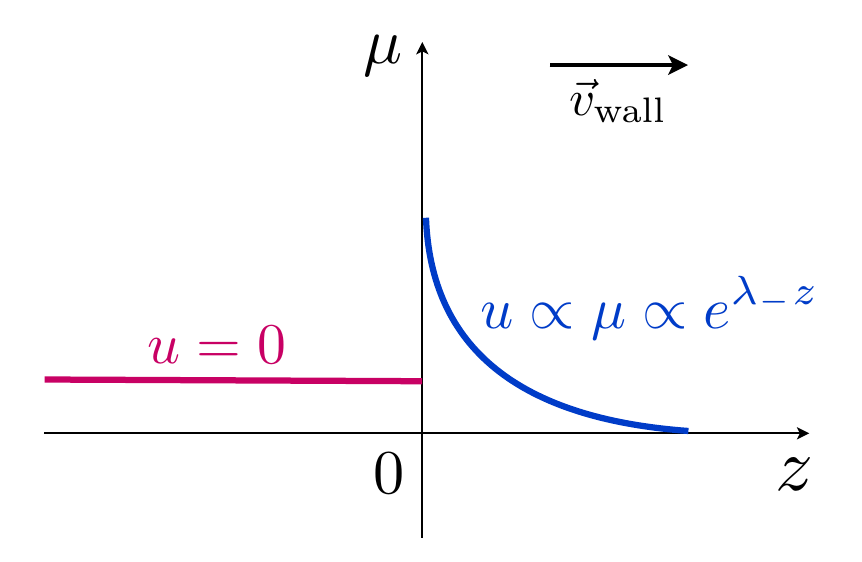}}
\caption{Sketch of the transport system~\eqref{eq:toytransport} solution for the chemical potential $\mu$ and the velocity perturbation $u$ across the phase transition wall, in the wall frame, with a point-like source at $z=0$, and assuming charge conservation behind the wall at $z<0$. $\vec v_{\text{wall}}$ shows the wall speed direction in the plasma frame. The forward diffusing part decays with a characteristic length $1/\lambda_-$ in front of the wall. The solution also demonstrates the trail of static ($u=0$) charge distribution left behind the source.}
\label{fig:transport}
\end{figure}

Let us now include a source of the asymmetry, using a simple toy model. Consider Eqs.~\eqref{eq:fluideq1} and~\eqref{eq:fluideq2} with a delta-function source,
\be
S_i=\delta(z)\,\hat s_i,
\ee
and approximately constant masses. The resulting system can be written as
\bea\label{eq:toytransport}
\left(\begin{matrix} 
|D_1| & 1 \\ 
D_2 & v_{\text w}
\end{matrix} \right)
\left(\begin{matrix} 
\mu' \\ 
u'
\end{matrix} \right)
- 
\left(\begin{matrix} 
\Gamma_\mu  & 0 \\ 
v_{\text{w}} \Gamma_\mu & \Gamma_u
\end{matrix} \right)
\left(\begin{matrix} 
\mu \\ 
u
\end{matrix} \right)
=
\left(\begin{matrix} 
S_1 \\ 
-S_2
\end{matrix} \right).
\eea
We take $\Gamma_\mu=0$ behind the wall ($z<0$), corresponding to exact conservation there, and for simplicity use the same velocity-damping rate $\Gamma_u$ on both sides. In the small-$v_{\text w}$ limit the analytic solution is (see Ref.~\cite{Azzola:2026cwa} for details) 
\bea
\mu_{z<0} &\simeq& \frac{\hat s_2}{D_2} - \frac{\Gamma_u}{\Gamma_\mu} \frac{D_2 \hat s_1 + |D_1| \hat s_2}{D_2^{3/2}},   \label{eq:mutoy1} \\
\mu_{z>0} &\simeq& - \frac{\Gamma_u}{\Gamma_\mu} \frac{D_2 \hat s_1 + |D_1| \hat s_2}{D_2^{3/2}}  \exp[\lambda_- z],
\eea
where the forward-diffusing solutions decays with the rate $\lambda_- <0$:
\be
\lambda_- \simeq - \frac{\sqrt{\Gamma_u \Gamma_\mu}}{\sqrt{D_2}} - \frac{|D_1| \Gamma_u}{2 D_2}.
\ee
Here $\Gamma_\mu$ denotes the charge-violating rate in front of the wall. The important lesson is that, as long as either $\hat s_1$ or $\hat s_2$ is nonzero, the wall leaves behind a nonzero constant chemical potential in the region where the corresponding charge is conserved.
This behavior is schematically shown in Fig.~\ref{fig:transport}. In the following section we will demonstrate that the results of the full numerical solution indeed follow the predicted behavior of the toy model (see black line in Fig.~\ref{fig:evolwall}).

\subsection{$X$-charge production}

We now apply this mechanism to the generation of the $U(1)_X$ charge carried by the inert Higgs doublet\footnote{Although a similar mechanism may also generate a sufficiently large $\tau_R$ asymmetry, we leave that possibility for future work.} $\Phi$ introduced in Section~\ref{s:chieq}. As in conventional EWBG, we take the primary source of CP violation to be a space-dependent complex phase of the top-quark mass,
\bea
m_t(z) &=& \frac{y_t}{\sqrt{2}} h(z) \left(1 + i \frac{S(z)}{\Lambda}\right),
\eea
where $h(z)$ is the Higgs profile and $S(z)$ is a real scalar background~\cite{Cline:2020jre}:
\bea\label{eq:profiles}
h(z) &=& \frac{v_n}{2}\left(1+ \tanh[ z/l_{\text{w}}] \right),\\
S(z) &=& \frac{w_n}{2}\left(1+ \tanh[ z/l_{\text{s}}] \right).
\eea
We treat these ingredients as an effective description, without committing to fully UV-complete them. Note that we could also use for example an SM-singlet $H H^\dagger/\Lambda$ combination instead of $S$.

The corresponding CP-violating sources entering the fluid equations~\eqref{eq:fluideq2} are~\cite{Cline:2020jre}:
\be\label{eq:SCPV}
(S^o_{h l})_t = - h\, v_{\text{w}} \gamma_{\text{w}} \left\{ (m_t^2 \theta_t')' Q_{l}^{8o} - (m_t^2)'m_t^2 \theta_t' Q^{9o}_{l} \right\},
\ee
where $Q_{l}^{8o}$ are momentum-averaged distribution functions, and $\theta_t$ is the top mass phase. The index $l=1,2$ labels the two moment equations. $h=\pm1$ distinguishes right and left-handed top quarks, and therefore such a source creates opposite excesses of $\mu_{t_L}$ and $\mu_{t_R}$. On its own, this combination does not feature any sufficiently good global charge, due to a sizeable top quark Yukawa and weak interactions\footnote{It is easy to see that such a combination features a non-zero hypercharge and weak isospin. In principle if such static charges were produced behind the wall they would result in a non-zero equilibrium baryon number. However, according to the studies of the direct EWPT~\cite{Khlebnikov:1992qs,Cohen:1992yh,Joyce:1994zn,Cline:1995di}, the initial hypercharge density in the unbroken phase drops considerably already at a distance $\sim 1/T$ from the wall. Hence we assume that also in our set-up the gauge charge density is quickly neutralized behind the wall.}, however it can itself act as a source for global $X$ charge production.

To demonstrate the mechanism we consider a simplistic model with an inert Higgs doublet $\Phi$ carrying the global charge $U(1)_X$. We assume that $U(1)_X$ is explicitly broken by the operator
\be\label{eq:phicpv}
\kappa\, S\, H^\dagger.\Phi  + h.c. ,
\ee
where $\kappa$ controls the size of the explicit $U(1)_X$ breaking. This operator can transfer the asymmetry generated in the Higgs sector to the inert doublet, provided the Higgs charge density is coherently sourced by the $t_L$ and $t_R$ asymmetries. Importantly, the explicit breaking is proportional to the local value of $S$, and is therefore effective only in front of the wall, where $S \neq 0$. Behind the wall, where $S$ vanishes, $U(1)_X$ is restored and the transferred $X$ asymmetry is conserved. We further assume that the physical excitations of $S$ are sufficiently heavy that virtual or on-shell $S$ exchange does not efficiently mediate chemical-potential redistribution in the plasma.

To model this transfer, we include the corresponding relaxation terms in the collision term $\delta C_1$, 
\be
\Gamma_{H\Phi}(\mu_{H} - \mu_{\Phi})\,, \qquad
\Gamma_{H\Phi} \simeq 0.5\,T\left(\frac{\kappa S(z)}{T^2}\right)^2,
\ee
where the rate is estimated conservatively, at the order-of-magnitude level, from the expressions in Ref.~\cite{Postma:2019scv}. For the remaining interactions, we take the inert doublet to have the same collision terms as the SM Higgs, except that Yukawa-mediated processes are absent. We also set $\langle \Phi \rangle = \langle H \rangle$ for simplicity.

\begin{figure}[t]
\center
{\includegraphics[width=0.7\textwidth]
{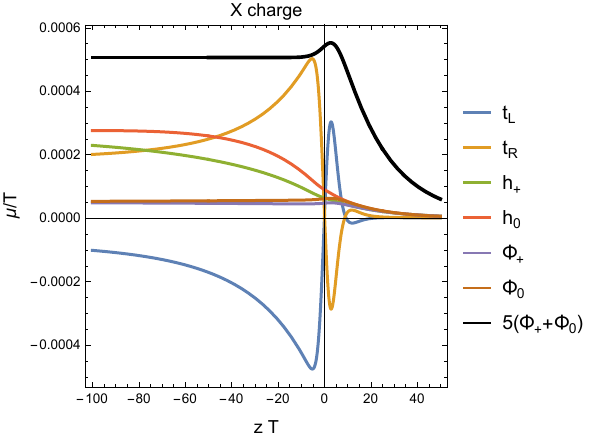}}
\caption{
Chemical potentials in the vicinity of the inverse electroweak wall. The wall is centered around $z=0$. The black curve shows the (multiplied by 5) combination $\mu_X=\mu_{\Phi_+}+\mu_{\Phi_0}$, that is conserved behind the wall, and determines the baryon asymmetry generated later through Eq.~\eqref{eq:nBXQ}.}
\label{fig:evolwall}
\end{figure}

To perform a numerical study we use the following benchmark that is similar to the parameter choice employed in Ref~\cite{Cline:2020jre} for the standard EWBG: 
\be\label{eq:params}
T= 120\text{ GeV},\; 
v_n = \kappa = 1.1 \, T,\;
w_n/\Lambda =1/3,\; 
l_{\text{w}} = l_{\text{s}}=5/T,\; 
v_{\text{w}}=0.4.
\ee
$w_n/\Lambda$ defines the size of CPV phase of the top mass; $v_n$ affects the CPV strength through $h(z)$ dependence, and also the $X$-charge violation strength; $l_{\text{w}}, l_{\text{s}}$ set the typical scales of the wall profile that affects all the transport dynamics, and the CPV generation; $l_{\text{w}}$ and $l_{\text{s}}$ are required to be $\gg 1/T$ for the validity of the gradient expansion underlying our approach.

Note that we could easily change the temperature to a higher value, and rescale other dimensionful parameters $\Lambda, v_n, w_n, \kappa, 1/l_{\text{w}}, 1/l_{\text{s}}$ by the same factor to obtain almost the same results (up to running effects).

The numerical asymmetry profiles are shown in Fig.\ref{fig:evolwall}~\footnote{Note that the apparently conserved larger asymmetries in other species than $\Phi$ at $z<0$ are partly an artifact: we did not include the gauge charge screening in our computation, and hence the predicted non-zero gauge charges support some asymmetries behind the wall. These asymmetries in the unbroken phase don't source the $X$ charge, and therefore we expect our predictions to not be strongly affected by this effect.}. The corresponding charge density $\mu_{X}/T \sim 10^{-4}$, induced by the wall can be combined with the $X/B$ ratio of Eq.~\eqref{eq:nBXQ} to obtain the resulting baryon-to-entropy ratio
\be
\eta^{(X)} = n_B \left(\frac{2 \pi^2}{45} g_* T^3 \right)^{-1} \sim \, 10^{-8},
\ee
to be compared with the observed value $10^{-10}$. This result was derived assuming that the top quarks produce a coherent superposition of mass eigenstates composing $H$, however if the mass splitting is large enough, the $\Phi$ mass eigenstate can be produced directly, breaking the coherence from the start, and requiring a more detailed analysis which we leave for future work.

Although we have not performed a detailed study of the wall-velocity dependence of the generated baryon asymmetry, the suppression of the CP-violating force at large wall velocities suggests that the walls in the inverse EWPT should not be too fast~\cite{Cline:2020jre,Dorsch:2021ubz}. Consequently, in the case of a global inverse EWPT proceeding via bubble nucleation, it may be necessary to introduce additional friction sources to slow down the phase-transition walls~\cite{Barni:2024lkj,Long:2025qoh}.

\section{Discussion}
\label{s:disc}

We have proposed a mechanism for baryogenesis in which weak sphalerons generate a nonzero equilibrium baryon number in the presence of an approximately conserved global charge produced during a cosmological phase transition. Our benchmark realization uses an inverse electroweak phase transition to generate an $X$ asymmetry carried by an inert Higgs doublet. Within this setup, we find that the observed baryon asymmetry can be reproduced at the level of order of magnitude for generic parameter choices.

The key point is that the mechanism does not rely on the standard electroweak-baryogenesis picture of diffusion in front of a wall followed by immediate sphaleron conversion. Instead, the wall deposits a conserved charge behind it, and weak sphalerons subsequently drive the plasma toward an equilibrium state with nonzero baryon number. In this sense, the baryon asymmetry is controlled primarily by the existence of the conserved charge and by the duration of the sphaleron-active phase, rather than directly by the weak sphaleron rate itself.

Although we focused on an inverse electroweak phase transition, the essential requirement is more general: the region behind the wall must support unsuppressed sphaleron processes, while the deposited charge remains approximately conserved there. A fully restored electroweak phase is therefore not strictly necessary; it is sufficient that $h/T\lesssim 1$ in the region behind the wall, so that sphalerons remain in equilibrium. Conversely, there is no strict requirement that $h/T \gtrsim 1$ in the broken phase, as is the case for standard EWBG. These relaxed conditions suggest that the mechanism can be extended to a wider class of phase transitions and defect-based realizations.

Our results also point to a phenomenology that is qualitatively different from conventional electroweak baryogenesis. The scenario requires new physics that provides CP violation at the wall, an altered phase-transition history, and an additional approximately conserved charge such as $X$. In concrete realizations this motivates electroweakly charged states carrying $X$, possibly with lifetimes long enough to protect the generated baryon asymmetry through the direct electroweak transition. Depending on the model, this may lead to collider signatures of exotic electroweak multiplets and to gravitational-wave signals associated with phase transitions at temperatures and with dynamics different from those usually considered in standard EWBG.

A particularly attractive feature is that the CP-violating sector need only be active during the inverse transition. Its characteristic mass scale can therefore be parametrically higher than in conventional electroweak baryogenesis, potentially weakening the usual tension with electric dipole moment bounds~\cite{ACME:2018yjb}. We thus view inverse electroweak baryogenesis as a qualitatively distinct framework, in which conserved charges, plasma neutrality, and sphaleron equilibration combine to generate the baryon asymmetry of the Universe.

\vspace{0.3cm}
{\bf Acknowledgements}

We are grateful to Thomas Konstandin for useful comments on the manuscript. 
This work was partially supported by the Collaborative Research Center SFB1258, the Munich Institute for Astro-, Particle and BioPhysics (MIAPbP), the Excellence Cluster ORIGINS, which is funded by the Deutsche Forschungsgemeinschaft (DFG, German Research Foundation) under Germany’s Excellence Strategy – EXC-2094-390783311. 

\vspace{1.3cm}

\appendix

\section{Chemical Potential and Number Density}
\label{s:appendix}

For a species $i$ in kinetic equilibrium with chemical potential $\mu_i$, 
\be
f(E)=\frac{1}{e^{(E-\mu_i)/T}\pm 1},\qquad 
\bar f(E)=\frac{1}{e^{(E+\mu_i)/T}\pm 1}.
\ee
with the upper (lower) sign for fermions (bosons).
The particle--antiparticle asymmetry is
\be
\delta n_i \equiv n_i-\bar n_i
= g_i\!\int\!\frac{d^3p}{(2\pi)^3}\Big[f(E,\mu_i)-f(E,-\mu_i)\Big].
\ee
For $|\mu_i|\ll T$,
$f(E,\mu)-f(E,-\mu)=2\mu\,(\partial f/\partial\mu)_{\mu=0}+\mathcal O(\mu^3)$ with
$\partial f/\partial\mu = \frac{1}{T} f(1\pm f)$, hence
\be
\delta n_i = \chi_i\,\mu_i,\qquad
\chi_i \equiv \frac{2g_i}{T}\!\int\!\frac{d^3p}{(2\pi)^3} f_0(E)\big(1\pm f_0(E)\big),
\ee
where $f_0(E)=1/(e^{E/T}\pm 1)$. In the relativistic limit $m_i\ll T$,
\be
\chi_i=\frac{g_iT^2}{6}\ \text{(Weyl fermion)},\qquad 
\chi_i=\frac{g_iT^2}{3}\ \text{(complex scalar)}.
\ee
We also use $(A,B)\equiv A^T\chi B=\sum_i A_i\chi_i B_i$.


\bibliographystyle{JHEP} 
\bibliography{biblio} 

\providecommand{\href}[2]{#2}\begingroup\raggedright\begin{thebibliography}{10}

\bibitem{Shaposhnikov:1987tw}
M.~E. Shaposhnikov, \emph{{Baryon Asymmetry of the Universe in Standard
  Electroweak Theory}},
  \href{https://doi.org/10.1016/0550-3213(87)90127-1}{\emph{Nucl. Phys. B}
  {\bfseries 287} (1987) 757--775}.

\bibitem{Cohen:1990it}
A.~G. Cohen, D.~B. Kaplan and A.~E. Nelson, \emph{{Baryogenesis at the weak
  phase transition}},
  \href{https://doi.org/10.1016/0550-3213(91)90395-E}{\emph{Nucl. Phys. B}
  {\bfseries 349} (1991) 727--742}.

\bibitem{Morrissey:2012db}
D.~E. Morrissey and M.~J. Ramsey-Musolf, \emph{{Electroweak baryogenesis}},
  \href{https://doi.org/10.1088/1367-2630/14/12/125003}{\emph{New J. Phys.}
  {\bfseries 14} (2012) 125003},
  [\href{https://arxiv.org/abs/1206.2942}{{\ttfamily 1206.2942}}].

\bibitem{Cline:2006ts}
J.~M. Cline, \emph{{Baryogenesis}},  in \emph{{Les Houches Summer School -
  Session 86: Particle Physics and Cosmology: The Fabric of Spacetime}}, 9,
  2006, \href{https://arxiv.org/abs/hep-ph/0609145}{{\ttfamily
  hep-ph/0609145}}.

\bibitem{Bhusal:2025lvm}
N.~Bhusal, S.~Blasi, M.~Cataldi, A.~Chatrchyan, M.~Gorghetto and G.~Servant,
  \emph{{Standard Model Baryon Number Violation at Zero Temperature from Higgs
  Bubble Collisions}},  \href{https://arxiv.org/abs/2508.21825}{{\ttfamily
  2508.21825}}.

\bibitem{Shakya:2025mdh}
B.~Shakya, \emph{{Did our Universe Tunnel out of the Wrong Higgs Vacuum?}},
  \href{https://arxiv.org/abs/2511.08843}{{\ttfamily 2511.08843}}.

\bibitem{Espinosa:2004pn}
J.~R. Espinosa, M.~Losada and A.~Riotto, \emph{{Symmetry nonrestoration at high
  temperature in little Higgs models}},
  \href{https://doi.org/10.1103/PhysRevD.72.043520}{\emph{Phys. Rev. D}
  {\bfseries 72} (2005) 043520},
  [\href{https://arxiv.org/abs/hep-ph/0409070}{{\ttfamily hep-ph/0409070}}].

\bibitem{DiLuzio:2019wsw}
L.~Di~Luzio, M.~Redi, A.~Strumia and D.~Teresi, \emph{{Coset Cosmology}},
  \href{https://doi.org/10.1007/JHEP06(2019)110}{\emph{JHEP} {\bfseries 06}
  (2019) 110}, [\href{https://arxiv.org/abs/1902.05933}{{\ttfamily
  1902.05933}}].

\bibitem{Glioti:2018roy}
A.~Glioti, R.~Rattazzi and L.~Vecchi, \emph{{Electroweak Baryogenesis above the
  Electroweak Scale}},
  \href{https://doi.org/10.1007/JHEP04(2019)027}{\emph{JHEP} {\bfseries 04}
  (2019) 027}, [\href{https://arxiv.org/abs/1811.11740}{{\ttfamily
  1811.11740}}].

\bibitem{Matsedonskyi:2022btb}
O.~Matsedonskyi, J.~Unwin and Q.~Wang, \emph{{Towards TeV-scale supersymmetric
  electroweak baryogenesis}},
  \href{https://doi.org/10.1007/JHEP02(2023)198}{\emph{JHEP} {\bfseries 02}
  (2023) 198}, [\href{https://arxiv.org/abs/2211.09147}{{\ttfamily
  2211.09147}}].

\bibitem{Baldes:2018nel}
I.~Baldes and G.~Servant, \emph{{High scale electroweak phase transition:
  baryogenesis {\textbackslash}{\&} symmetry non-restoration}},
  \href{https://doi.org/10.1007/JHEP10(2018)053}{\emph{JHEP} {\bfseries 10}
  (2018) 053}, [\href{https://arxiv.org/abs/1807.08770}{{\ttfamily
  1807.08770}}].

\bibitem{Matsedonskyi:2020kuy}
O.~Matsedonskyi, \emph{{High-Temperature Electroweak Symmetry Breaking by SM
  Twins}}, \href{https://doi.org/10.1007/JHEP04(2021)036}{\emph{JHEP}
  {\bfseries 04} (2021) 036},
  [\href{https://arxiv.org/abs/2008.13725}{{\ttfamily 2008.13725}}].

\bibitem{Matsedonskyi:2020mlz}
O.~Matsedonskyi and G.~Servant, \emph{{High-Temperature Electroweak Symmetry
  Non-Restoration from New Fermions and Implications for Baryogenesis}},
  \href{https://doi.org/10.1007/JHEP09(2020)012}{\emph{JHEP} {\bfseries 09}
  (2020) 012}, [\href{https://arxiv.org/abs/2002.05174}{{\ttfamily
  2002.05174}}].

\bibitem{Badziak:2025fdp}
M.~Badziak, K.~Harigaya and I.~Nalkecz, \emph{{Electroweak symmetry
  non-restoration and suppressed dark radiation in Supersymmetric Twin Higgs
  model}},  \href{https://arxiv.org/abs/2508.15894}{{\ttfamily 2508.15894}}.

\bibitem{Biekotter:2022kgf}
T.~Biek{\"o}tter, S.~Heinemeyer, J.~M. No, M.~O. Olea-Romacho and G.~Weiglein,
  \emph{{The trap in the early Universe: impact on the interplay between
  gravitational waves and LHC physics in the 2HDM}},
  \href{https://doi.org/10.1088/1475-7516/2023/03/031}{\emph{JCAP} {\bfseries
  03} (2023) 031}, [\href{https://arxiv.org/abs/2208.14466}{{\ttfamily
  2208.14466}}].

\bibitem{Inoue:2015pza}
S.~Inoue, G.~Ovanesyan and M.~J. Ramsey-Musolf, \emph{{Two-Step Electroweak
  Baryogenesis}}, \href{https://doi.org/10.1103/PhysRevD.93.015013}{\emph{Phys.
  Rev. D} {\bfseries 93} (2016) 015013},
  [\href{https://arxiv.org/abs/1508.05404}{{\ttfamily 1508.05404}}].

\bibitem{Biekotter:2021ysx}
T.~Biek\"otter, S.~Heinemeyer, J.~M. No, M.~O. Olea and G.~Weiglein,
  \emph{{Fate of electroweak symmetry in the early Universe: Non-restoration
  and trapped vacua in the N2HDM}},
  \href{https://doi.org/10.1088/1475-7516/2021/06/018}{\emph{JCAP} {\bfseries
  06} (2021) 018}, [\href{https://arxiv.org/abs/2103.12707}{{\ttfamily
  2103.12707}}].

\bibitem{Aoki:2023lbz}
M.~Aoki, L.~Biermann, C.~Borschensky, I.~P. Ivanov, M.~M\"uhlleitner and
  H.~Shibuya, \emph{{Intermediate charge-breaking phases and symmetry
  non-restoration in the 2-Higgs-Doublet Model}},
  \href{https://doi.org/10.1007/JHEP02(2024)232}{\emph{JHEP} {\bfseries 02}
  (2024) 232}, [\href{https://arxiv.org/abs/2308.04141}{{\ttfamily
  2308.04141}}].

\bibitem{Ai:2025vfi}
W.-Y. Ai, P.~Huang and K.-P. Xie, \emph{{When inverse seesaw meets inverse
  electroweak phase transition: a novel path to leptogenesis}},
  \href{https://doi.org/10.1007/JHEP02(2026)256}{\emph{JHEP} {\bfseries 02}
  (2026) 256}, [\href{https://arxiv.org/abs/2510.09000}{{\ttfamily
  2510.09000}}].

\bibitem{Barni:2025ced}
G.~Barni and A.~Tesi, \emph{{Super-heated first order phase transitions}},
  \href{https://doi.org/10.1007/JHEP03(2026)188}{\emph{JHEP} {\bfseries 03}
  (2026) 188}, [\href{https://arxiv.org/abs/2508.08362}{{\ttfamily
  2508.08362}}].

\bibitem{Barni:2024lkj}
G.~Barni, S.~Blasi and M.~Vanvlasselaer, \emph{{The hydrodynamics of inverse
  phase transitions}},
  \href{https://doi.org/10.1088/1475-7516/2024/10/042}{\emph{JCAP} {\bfseries
  10} (2024) 042}, [\href{https://arxiv.org/abs/2406.01596}{{\ttfamily
  2406.01596}}].

\bibitem{Badziak:2022ltm}
M.~Badziak and I.~Nalkecz, \emph{{First-order phase transitions in Twin Higgs
  models}}, \href{https://doi.org/10.1007/JHEP02(2023)185}{\emph{JHEP}
  {\bfseries 02} (2023) 185},
  [\href{https://arxiv.org/abs/2212.09776}{{\ttfamily 2212.09776}}].

\bibitem{Azzola:2026cwa}
J.~Azzola, O.~Matsedonskyi and A.~Weiler, \emph{{Baryon Asymmetry from
  Electroweak-Symmetric Domain Walls}},
  \href{https://arxiv.org/abs/2604.16603}{{\ttfamily 2604.16603}}.

\bibitem{Brandenberger:1994mq}
R.~H. Brandenberger, A.-C. Davis, T.~Prokopec and M.~Trodden, \emph{{Local and
  nonlocal defect mediated electroweak baryogenesis}},
  \href{https://doi.org/10.1103/PhysRevD.53.4257}{\emph{Phys. Rev. D}
  {\bfseries 53} (1996) 4257--4266},
  [\href{https://arxiv.org/abs/hep-ph/9409281}{{\ttfamily hep-ph/9409281}}].

\bibitem{Abel:1995uc}
S.~A. Abel and P.~L. White, \emph{{Baryogenesis from domain walls in the
  next-to-minimal supersymmetric standard model}},
  \href{https://doi.org/10.1103/PhysRevD.52.4371}{\emph{Phys. Rev. D}
  {\bfseries 52} (1995) 4371--4379},
  [\href{https://arxiv.org/abs/hep-ph/9505241}{{\ttfamily hep-ph/9505241}}].

\bibitem{Brandenberger:2005bx}
R.~H. Brandenberger, W.~Kelly and M.~Yamaguchi, \emph{{Electroweak baryogenesis
  with embedded domain walls}},
  \href{https://doi.org/10.1143/PTP.117.823}{\emph{Prog. Theor. Phys.}
  {\bfseries 117} (2007) 823--834},
  [\href{https://arxiv.org/abs/hep-ph/0503211}{{\ttfamily hep-ph/0503211}}].

\bibitem{Bai:2021xyf}
Y.~Bai, J.~Berger, M.~Korwar and N.~Orlofsky, \emph{{Catalyzed baryogenesis}},
  \href{https://doi.org/10.1007/JHEP10(2021)147}{\emph{JHEP} {\bfseries 10}
  (2021) 147}, [\href{https://arxiv.org/abs/2106.12589}{{\ttfamily
  2106.12589}}].

\bibitem{Schroder:2024gsi}
T.~Schr\"oder and R.~Brandenberger, \emph{{Embedded domain walls and
  electroweak baryogenesis}},
  \href{https://doi.org/10.1103/PhysRevD.110.043516}{\emph{Phys. Rev. D}
  {\bfseries 110} (2024) 043516},
  [\href{https://arxiv.org/abs/2404.13035}{{\ttfamily 2404.13035}}].

\bibitem{Azzola:2024pzq}
J.~Azzola, O.~Matsedonskyi and A.~Weiler, \emph{{Minimal electroweak
  baryogenesis via domain walls}},
  \href{https://doi.org/10.1007/JHEP04(2025)103}{\emph{JHEP} {\bfseries 04}
  (2025) 103}, [\href{https://arxiv.org/abs/2412.10495}{{\ttfamily
  2412.10495}}].

\bibitem{Sassi:2024cyb}
M.~Y. Sassi and G.~Moortgat-Pick, \emph{{Electroweak symmetry restoration in
  the N2HDM via domain walls}},
  \href{https://doi.org/10.1007/JHEP06(2025)072}{\emph{JHEP} {\bfseries 06}
  (2025) 072}, [\href{https://arxiv.org/abs/2407.14468}{{\ttfamily
  2407.14468}}].

\bibitem{Sakharov:1967dj}
A.~D. Sakharov, \emph{{Violation of CP Invariance, C asymmetry, and baryon
  asymmetry of the universe}},
  \href{https://doi.org/10.1070/PU1991v034n05ABEH002497}{\emph{Pisma Zh. Eksp.
  Teor. Fiz.} {\bfseries 5} (1967) 32--35}.

\bibitem{Cline:1993vv}
J.~M. Cline, K.~Kainulainen and K.~A. Olive, \emph{{On the erasure and
  regeneration of the primordial baryon asymmetry by sphalerons}},
  \href{https://doi.org/10.1103/PhysRevLett.71.2372}{\emph{Phys. Rev. Lett.}
  {\bfseries 71} (1993) 2372--2375},
  [\href{https://arxiv.org/abs/hep-ph/9304321}{{\ttfamily hep-ph/9304321}}].

\bibitem{Antaramian:1993nt}
A.~Antaramian, L.~J. Hall and A.~Rasin, \emph{{Hypercharge and the cosmological
  baryon asymmetry}},
  \href{https://doi.org/10.1103/PhysRevD.49.3881}{\emph{Phys. Rev. D}
  {\bfseries 49} (1994) 3881--3885},
  [\href{https://arxiv.org/abs/hep-ph/9311279}{{\ttfamily hep-ph/9311279}}].

\bibitem{Dick:1999je}
K.~Dick, M.~Lindner, M.~Ratz and D.~Wright, \emph{{Leptogenesis with Dirac
  neutrinos}}, \href{https://doi.org/10.1103/PhysRevLett.84.4039}{\emph{Phys.
  Rev. Lett.} {\bfseries 84} (2000) 4039--4042},
  [\href{https://arxiv.org/abs/hep-ph/9907562}{{\ttfamily hep-ph/9907562}}].

\bibitem{Joyce:1994fu}
M.~Joyce, T.~Prokopec and N.~Turok, \emph{{Electroweak baryogenesis from a
  classical force}},
  \href{https://doi.org/10.1103/PhysRevLett.75.1695}{\emph{Phys. Rev. Lett.}
  {\bfseries 75} (1995) 1695--1698},
  [\href{https://arxiv.org/abs/hep-ph/9408339}{{\ttfamily hep-ph/9408339}}].

\bibitem{Joyce:1994zt}
M.~Joyce, T.~Prokopec and N.~Turok, \emph{{Nonlocal electroweak baryogenesis.
  Part 2: The Classical regime}},
  \href{https://doi.org/10.1103/PhysRevD.53.2958}{\emph{Phys. Rev. D}
  {\bfseries 53} (1996) 2958--2980},
  [\href{https://arxiv.org/abs/hep-ph/9410282}{{\ttfamily hep-ph/9410282}}].

\bibitem{Cline:2000kb}
J.~M. Cline and K.~Kainulainen, \emph{{A New source for electroweak
  baryogenesis in the MSSM}},
  \href{https://doi.org/10.1103/PhysRevLett.85.5519}{\emph{Phys. Rev. Lett.}
  {\bfseries 85} (2000) 5519--5522},
  [\href{https://arxiv.org/abs/hep-ph/0002272}{{\ttfamily hep-ph/0002272}}].

\bibitem{Cline:2020jre}
J.~M. Cline and K.~Kainulainen, \emph{{Electroweak baryogenesis at high bubble
  wall velocities}},
  \href{https://doi.org/10.1103/PhysRevD.101.063525}{\emph{Phys. Rev. D}
  {\bfseries 101} (2020) 063525},
  [\href{https://arxiv.org/abs/2001.00568}{{\ttfamily 2001.00568}}].

\bibitem{Harvey:1990qw}
J.~A. Harvey and M.~S. Turner, \emph{{Cosmological Baryon and Lepton Number in
  the Presence of Electroweak Fermion Number Violation}},
  \href{https://doi.org/10.1103/PhysRevD.42.3344}{\emph{Phys. Rev. D}
  {\bfseries 42} (1990) 3344--3349}.

\bibitem{Cline:2021dkf}
J.~M. Cline and B.~Laurent, \emph{{Electroweak baryogenesis from light fermion
  sources: A critical study}},
  \href{https://doi.org/10.1103/PhysRevD.104.083507}{\emph{Phys. Rev. D}
  {\bfseries 104} (2021) 083507},
  [\href{https://arxiv.org/abs/2108.04249}{{\ttfamily 2108.04249}}].

\bibitem{Barni:2025ifb}
G.~Barni, \emph{{Electroweak Baryogenesis with BARYONET: a self-contained
  review of the WKB approach}},
  \href{https://arxiv.org/abs/2510.21915}{{\ttfamily 2510.21915}}.

\bibitem{Li:2025kyo}
X.-X. Li, M.~J. Ramsey-Musolf, T.~V.~I. Tenkanen and Y.~Wu, \emph{{An Effective
  Sphaleron Awakens}},  \href{https://arxiv.org/abs/2506.01585}{{\ttfamily
  2506.01585}}.

\bibitem{Bodeker:2019ajh}
D.~B{\"o}deker and D.~Schr{\"o}der, \emph{{Equilibration of right-handed
  electrons}}, \href{https://doi.org/10.1088/1475-7516/2019/05/010}{\emph{JCAP}
  {\bfseries 05} (2019) 010},
  [\href{https://arxiv.org/abs/1902.07220}{{\ttfamily 1902.07220}}].

\bibitem{Khlebnikov:1992qs}
S.~Y. Khlebnikov, \emph{{Nonequilibrium charge transport at a first order phase
  transition}}, \href{https://doi.org/10.1016/0370-2693(93)91349-R}{\emph{Phys.
  Lett. B} {\bfseries 300} (1993) 376--380}.

\bibitem{Cohen:1992yh}
A.~G. Cohen, D.~B. Kaplan and A.~E. Nelson, \emph{{Debye screening and
  baryogenesis during the electroweak phase transition}},
  \href{https://doi.org/10.1016/0370-2693(92)91640-U}{\emph{Phys. Lett. B}
  {\bfseries 294} (1992) 57--62},
  [\href{https://arxiv.org/abs/hep-ph/9206214}{{\ttfamily hep-ph/9206214}}].

\bibitem{Joyce:1994zn}
M.~Joyce, T.~Prokopec and N.~Turok, \emph{{Nonlocal electroweak baryogenesis.
  Part 1: Thin wall regime}},
  \href{https://doi.org/10.1103/PhysRevD.53.2930}{\emph{Phys. Rev. D}
  {\bfseries 53} (1996) 2930--2957},
  [\href{https://arxiv.org/abs/hep-ph/9410281}{{\ttfamily hep-ph/9410281}}].

\bibitem{Cline:1995di}
J.~M. Cline and K.~Kainulainen, \emph{{Diffusion and Debye screening near
  expanding domain walls}},
  \href{https://doi.org/10.1016/0370-2693(95)00791-I}{\emph{Phys. Lett. B}
  {\bfseries 356} (1995) 19--25},
  [\href{https://arxiv.org/abs/hep-ph/9506285}{{\ttfamily hep-ph/9506285}}].

\bibitem{Postma:2019scv}
M.~Postma and J.~van~de Vis, \emph{{Source terms for electroweak baryogenesis
  in the vev-insertion approximation beyond leading order}},
  \href{https://doi.org/10.1007/JHEP02(2020)090}{\emph{JHEP} {\bfseries 02}
  (2020) 090}, [\href{https://arxiv.org/abs/1910.11794}{{\ttfamily
  1910.11794}}].

\bibitem{Dorsch:2021ubz}
G.~C. Dorsch, S.~J. Huber and T.~Konstandin, \emph{{On the wall velocity
  dependence of electroweak baryogenesis}},
  \href{https://doi.org/10.1088/1475-7516/2021/08/020}{\emph{JCAP} {\bfseries
  08} (2021) 020}, [\href{https://arxiv.org/abs/2106.06547}{{\ttfamily
  2106.06547}}].

\bibitem{Long:2025qoh}
A.~J. Long, B.~Shakya and J.~A. Ziegler, \emph{{Bubble Friction in
  Symmetry-Restoring Transitions}},
  \href{https://arxiv.org/abs/2511.10415}{{\ttfamily 2511.10415}}.

\bibitem{ACME:2018yjb}
{\scshape ACME} collaboration, V.~Andreev et~al., \emph{{Improved limit on the
  electric dipole moment of the electron}},
  \href{https://doi.org/10.1038/s41586-018-0599-8}{\emph{Nature} {\bfseries
  562} (2018) 355--360}.

\end{thebibliography}\endgroup

\end{document}